\documentclass[entropy,article,submit,moreauthors,pdftex,10pt,a4paper]{Definitions/mdpi} 

\usepackage{amsmath}
\usepackage{graphicx}
\usepackage{subfigure}
\usepackage{amsthm}
\usepackage{verbatim}
\usepackage{dcolumn}
\usepackage{bm}
\usepackage{epsf}
\usepackage{color}
\usepackage{dsfont}
\usepackage{amssymb}
\usepackage{stmaryrd}
\newcommand\myeqref[1]{
	Eq. (\textup{\ref{#1}})
}

\newcommand{\bla}{bla\\bla\\bla\\bla\\bla}

\newcommand{\mc}[1]{\mathcal{#1}}

\firstpage{1} 
\pubvolume{xx}
\issuenum{1}
\articlenumber{5}
\pubyear{2018}
\copyrightyear{2018}
\history{Received: date; Accepted: date; Published: date}

\Title{Environment-Assisted Shortcuts to Adiabaticity}

\Author{Akram Touil $^{1,*}$\orcidB{} and Sebastian Deffner $^{1,2}$\orcidC{}}
\address{%
	$^{1}$ \quad Department of Physics, University of Maryland, Baltimore County, Baltimore, MD 21250, USA\\
	$^{2}$ \quad Instituto de F\'isica `Gleb Wataghin', Universidade Estadual de Campinas, 13083-859, Campinas, S\~{a}o Paulo, Brazil}

\corres{Correspondence: akramt1@umbc.edu}

\abstract{Envariance is a symmetry exhibited by correlated quantum systems. Inspired by this ``quantum fact of life,'' we propose a novel method for shortcuts to adiabaticity which enables the system to evolve through the adiabatic manifold at all times, solely by controlling the environment. As main results, we construct the unique form of the driving on the environment that enables such dynamics, for a family of composite states of arbitrary dimension. We compare the cost of this environment-assisted technique with that of counterdiabatic driving, and we illustrate our results for a two-qubit model.}

\preto{\abstractkeywords}{\nolinenumbers}
\begin{document}

\section{Introduction}

An essential step in the development of viable quantum technologies is to achieve precise control over quantum dynamics \cite{techno1,techno2}.  In many situations, optimal performance relies on the ability to create particular target states. However,  in dynamically reaching such states the quantum adiabatic theorem \cite{adtheorem} poses a formidable challenge, since finite-time driving inevitably causes parasitic excitations \cite{transitions1,transitions2,transitions3,transitions4}. Acknowledging and addressing this issue, the field of ``shortcuts to adiabaticity'' (STA) \cite{Torrontegui2013,review1,review2,review3} has developed a variety of techniques that permit to facilitate effectively adiabatic dynamics in finite time.

Recent years have seen an explosion of work on, for instance, counterdiabatic driving \cite{CD1,CD2,CD3,LMGsta,psta2,psta3,psta4,psta5}, the fast-forward method~\cite{fastforw1,fastforw2,fastforw3,fastforw4}, time-rescaling~\cite{resta1,resta2},  methods based on identifying the adiabatic invariant~\cite{invar1,invar2,invar3,invar4}, and even generalizations to classical dynamics \cite{csta1,csta2,Iram2020}. For comprehensive reviews of the various techniques  we refer to the recent literature \cite{review1,review2,review3}.

Among these different paradigms, counterdiabatic driving (CD) stands out as it is the only method that forces evolution through the adiabatic manifold at all times.  However, experimentally realizing the CD method requires applying a complicated control field, which often involves non-local terms that are hard to implement in many-body systems~\cite{LMGsta,psta3}.  This may be particularly challenging if the system is not readily accessible, as for instance due to geometric restrictions of the experimental set-up. 

In the present paper, we propose an alternative method to achieve transitionless quantum driving by leveraging the system's (realistically) inevitable interaction with the environment.  Our novel paradigm is inspired by ``envariance,'' which is short for entanglement-assisted invariance.  Envariance is a symmetry of composite quantum systems, first described by Wojciech H. Zurek~\cite{born1}.  Consider a quantum state $|\psi_{\mc{SE}}\rangle$ that lives on a composite quantum universe comprising the system, $\mc{S}$, and its environment, $\mc{E}$. Then,  $|\psi_{\mc{SE}}\rangle$ is called envariant under a unitary map $u_{\mc{S}} \otimes \mathbb{I}_{\mc{E}}$ if and only if there exists another unitary $\mathbb{I}_{\mc{S}} \otimes u_{\mc{E}}$ acting on $\mc{E}$, such that the composite state remains unaltered after applying both maps, i.e., $\left(u_{\mc{S}} \otimes \mathbb{I}_{\mc{E}}\right)|\psi_{\mc{SE}}\rangle= |\phi_{\mc{SE}}\rangle$ and $\left(\mathbb{I}_{\mc{S}} \otimes u_{\mc{E}}\right)|\phi_{\mc{SE}}\rangle= |\psi_{\mc{SE}}\rangle$. In other words, the state is envariant if the action of a unitary on $\mc{S}$ can be inverted by applying a unitary on $\mc{E}$. 

Envariance has been essential to derive Born's rule~\cite{born1,born2}, and in formulating a novel approach to the foundations of statistical mechanics~\cite{deffner2016foundations}. Moreover, experiments~\cite{envexp1,envexp2} showed that this inherent symmetry of composite quantum states is indeed a physical reality, or rather a ``quantum fact of life'' with no classical analog~\cite{born2}. Drawing inspiration from envariance, we develop a novel method for transitionless quantum driving. In the following, we will see that instead of inverting the action of a unitary on $\mc{S}$, we can suppress undesirable transitions in the energy eigenbasis of $\mc{S}$ by applying a control field on the environment $\mc{E}$.  In particular, we consider the unitary evolution of an ensemble of composite states $\{|\psi_{\mc{SE}}\rangle\}$, on a Hilbert space $\mc{H}_{\mc{S}}\otimes \mc{H}_{\mc{E}}$ of arbitrary dimension, and we determine the general analytic form of the time-dependent driving on $\mc{H}_{\mc{E}}$ which suppresses undesirable transitions in the system of interest $\mc{S}$. This general driving on the environment $\mc{E}$ guarantees that the system $\mc{S}$ evolves through the adiabatic manifold at all times. We dub this technique \textit{Environment-Assisted Shortcuts To Adiabaticity}, or ``EASTA'' for short.  In addition, we prove that the cost associated with the EASTA technique is exactly equal to that of counterdiabatic driving. We illustrate our results in a simple two-qubit model, where the system and the environment are each described by a single qubit. Finally, we conclude with discussing a few implications of our results in the general context of decoherence theory and Quantum Darwinism. 

\section{Counterdiabatic Driving}

We start by briefly reviewing counterdiabatic driving to establish notions and notations.  Consider a quantum system $\mc{S}$, in a Hilbert space $\mc{H}_{\mc{S}}$ of dimension $d_{\mc{S}}$, driven by the Hamiltonian $H_{0}(t)$ with instantaneous eigenvalues $\{E_{n}(t)\}_{n \in \llbracket 0, \ d_{\mc{S}}-1 \rrbracket}$ and eigenstates $\{|n(t) \rangle \}_{n \in \llbracket 0, \ d_{\mc{S}}-1 \rrbracket}$. For slowly varying $H_{0}(t)$, according to the quantum adiabatic theorem~\cite{adtheorem}, the driving of $\mc{S}$ is transitionless. In other words, if the system starts in the eigenstate $|n(0) \rangle$, at $t=0$, it evolves into the eigenstate $|n(t) \rangle$ at time $t$ (with a phase factor),
\begin{equation}
|\psi_n(t)\rangle\equiv U(t)|n(0) \rangle = e^{-\frac{i}{\hbar} \int_{0}^{t}  E_{n}(s) ds- \int_{0}^{t} \left\langle n \mid \partial_{s} n \right\rangle ds} |n(t) \rangle \equiv e^{-\frac{i}{\hbar} f_{n}(t)} |n(t) \rangle.
\end{equation}

For arbitrary driving $H_{0}(t)$, namely for driving rates larger than the typical energy gaps, the system undergoes transitions. However, it has been shown~\cite{CD1,CD2,CD3} that the addition of a counterdiabatic field $H_{\text{\tiny CD}}(t)$ forces the system to evolve through the adiabatic manifold. Using the total Hamiltonian
\begin{equation}
H = H_{0}(t) +H_{\text{\tiny CD}}(t) = H_{0}(t)+i \hbar \sum_{n}\left(\left|\partial_{t} n\right\rangle\left\langle n\left|-\left\langle n \mid \partial_{t} n\right\rangle\right| n\right\rangle\langle n|\right)\,,
\end{equation}
 the system evolves with the corresponding unitary $U_{\text{\tiny CD}}(t)= \sum_{n} |\psi_n(t)\rangle \langle n(0)|$ such that
\begin{equation}
	U_{\text{\tiny CD}}(t)|n(0) \rangle = e^{-\frac{i}{\hbar} f_{n}(t)} |n(t) \rangle.
\end{equation}
This evolution is exact no matter how fast the system is driven by the total Hamiltonian. However, the counterdiabatic driving (CD) method requires adding a complicated counterdiabatic field $H_{\text{\tiny CD}}(t)$ involving highly non-local terms that are hard to implement in a many-body set-up~\cite{LMGsta,psta3}. Constructing this counterdiabatic field requires determining the instantaneous eigenstates $\{|n(t) \rangle \}_{n \in \llbracket 0, \ d_{\mc{S}}-1 \rrbracket}$ of the time-dependent Hamiltonian $H_{0}(t)$.  Moreover,  changing the dynamics of the system of interest (i.e., adding the counterdiabatic field), requires direct access and control on $\mc{S}$. 

In the following, we will see how (at least) the second issue can be circumvented by relying on the environment $\mc{E}$ that inevitably couples to the system of interest. In particular, we  make use of the entanglement between system and environment to avoid any transitions in the system.  To this end, we construct the unique  driving of the environment $\mc{E}$ that counter-acts the transitions in $\mc{S}$.

\section{Open system dynamics and STA for mixed states}
\label{sec3}

We start by stating three crucial assumptions: (i) the joint state of the system $\mc{S}$ and the environment $\mc{E}$ is described by an initial wave function $| \psi_{\mc{SE}}(0) \rangle$ evolving unitarly according to the Schr{\"o}dinger equation; (ii) the environment's degrees of freedom do not interact with each other; (iii) the $\mc{S}$-$\mc{E}$ joint state belongs to the ensemble of \textit{singly branching states}~\cite{BKWHZ}. These branching states have the  general form,
\begin{equation}
| \psi_{\mc{SE}} \rangle=  \sum^{N-1}_{n=0} \sqrt{p_n} |n \rangle \bigotimes^{N_{\mc{E}}}_{l=1}|\mc{E}^{l}_n \rangle,
\label{branch1}
\end{equation}
where $p_n \in [0,\ 1]$ is the probability associated with the $n$th branch of the wave function, with orthonormal states $|n \rangle \in \mc{H}_{\mc{S}}$ and $\bigotimes^{N_{\mc{E}}}_{l=1} |\mc{E}^{l}_n \rangle \in \mc{H}_{\mc{E}}$. 

Without loss of generality we can further assume $\sqrt{p_n}=1/\sqrt{N}$ for all $n \in \llbracket 0, \ N-1 \rrbracket$, since if $\sqrt{p_n}\neq 1/\sqrt{N}$ we can always find an extended Hilbert space~\cite{born1,born2} such that the state $| \psi_{\mc{SE}} \rangle$ becomes even. Thus,  we can consider branching state $| \psi_{\mc{SE}} \rangle$ of the simpler form
\begin{equation}
| \psi_{\mc{SE}} \rangle=  \frac{1}{\sqrt{N}} \sum^{N-1}_{n=0}  |n \rangle \bigotimes^{N_{\mc{E}}}_{l=1}|\mc{E}^{l}_n \rangle.
\label{branch2}
\end{equation}
In the following, we will see that EASTA can actually only be facilitated for even states \eqref{branch2}\footnote{In Appendix~\ref{a}, we show that EASTA cannot be implemented for arbitrary probabilities (i.e., $(\exists \ n);  \ \sqrt{p_n} \neq 1/\sqrt{N}$).}.

\subsection{Two-level environment $\mc{E}$}

\begin{figure}[h!]
	\centering
	\subfigure[Counterdiabatic scheme for open system dynamics]{
		\includegraphics[width=.9\textwidth]{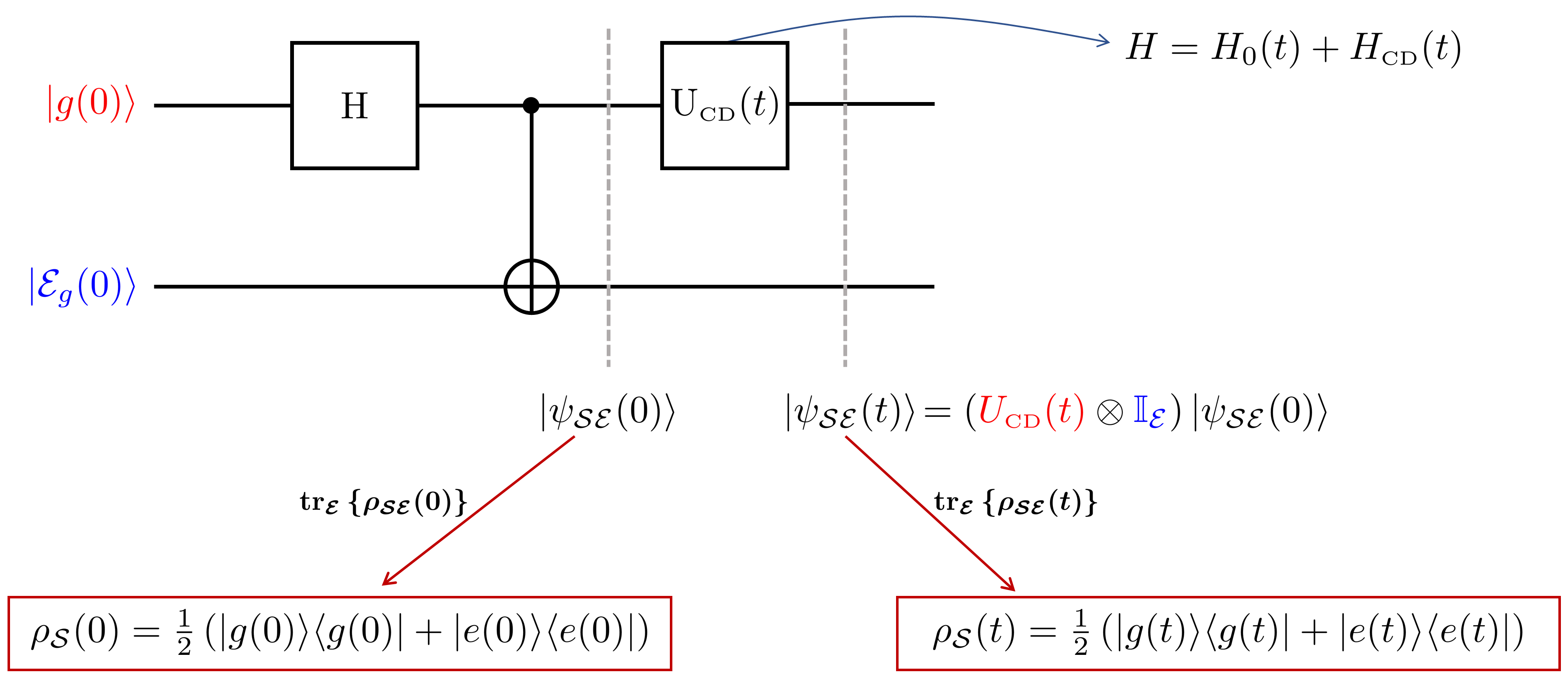}
	}
	\subfigure[Environment-assisted shortcut scheme]{
		\includegraphics[width=.9\textwidth]{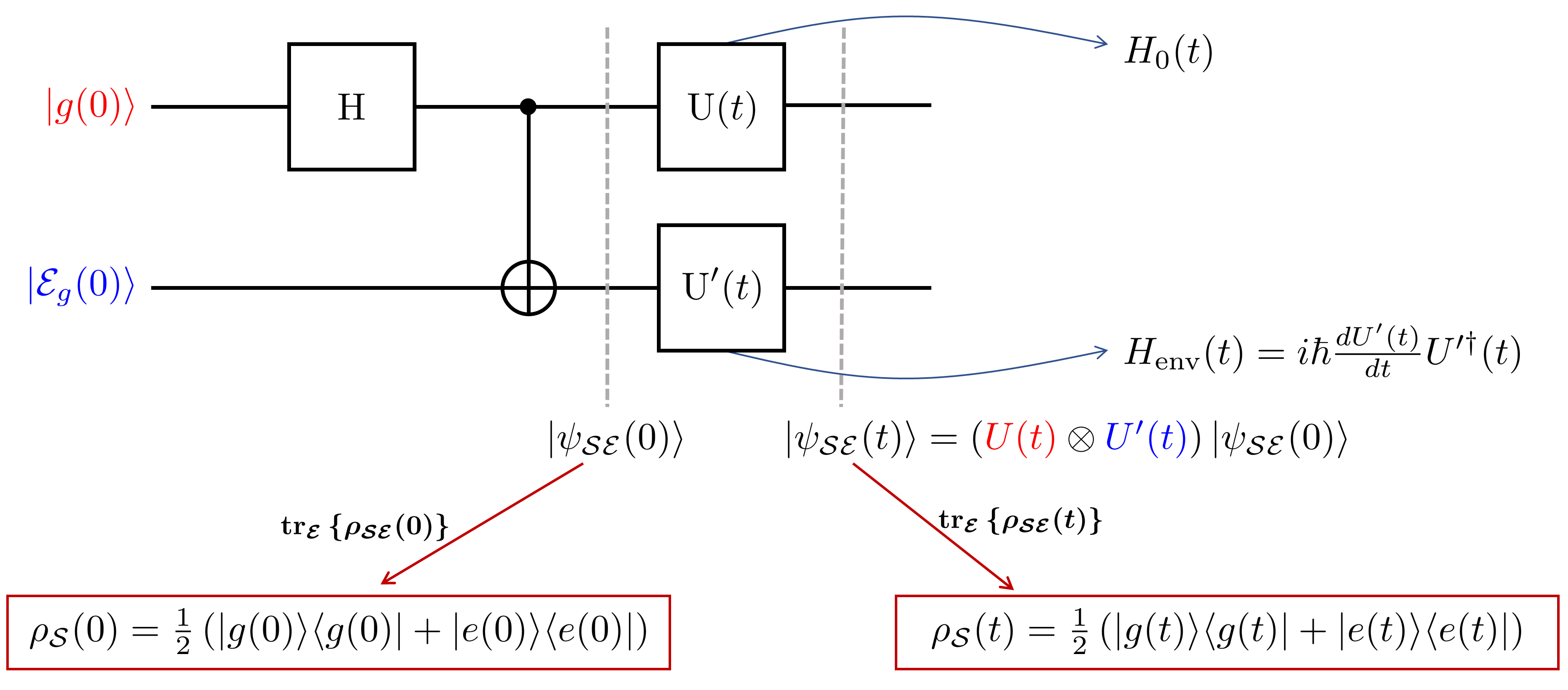}
	}
	\caption[]{\label{sta}Sketch of the two different schemes of applying STA in a branching state of the form presented in~\myeqref{twol}. In both panels, the state preparation involves a Hadamard gate (H) applied on $\mc{S}$, and coupling with the environment through a {\tt c-not} operation. In panel (a), we describe the ``usual'' counterdiabatic scheme. As shown in Section~\ref{sec3}, local driving on $\mc{E}$ suppresses any transitions of $\mc{S}$ in the instantaneous eigenbasis of $H_{0}(t)$. The latter scheme is illustrated in panel (b).}
\end{figure}

We start with the instructive case of a two-level environment, cf.  Fig.~\ref{sta}.  To this end, consider the branching state
\begin{equation}
| \psi_{\mc{SE}}(0) \rangle= \frac{1}{\sqrt{2}} |g (0) \rangle\otimes|\mc{E}_g (0) \rangle+\frac{1}{\sqrt{2}} |e (0) \rangle\otimes|\mc{E}_e (0) \rangle,
\label{twol}
\end{equation}
where the states $|\mc{E}_g (0) \rangle$ and $|\mc{E}_e (0) \rangle$ form a basis on the environment $\mc{E}$, and the states $|g (0) \rangle$ and $|e (0) \rangle$ represent the ground and excited states of $\mc{S}$ at $t=0$, respectively. 

It is then easy to see that there exists a unique unitary $U^{\prime}$ such that the system evolves through the adiabatic manifold in each branch of the wave function, 
\begin{equation}
(\exists! \ U^{\prime}); \ \ (U \otimes U^{\prime})| \psi_{\mc{SE}}(0) \rangle= (U_{\text{\tiny CD}} \otimes \mathds{I}_{\mc{E}})| \psi_{\mc{SE}}(0) \rangle.
\end{equation}
Starting from the above equality, we obtain
\begin{equation}
U |g (0) \rangle\otimes U^{\prime}|\mc{E}_g (0) \rangle + U |e (0) \rangle \otimes U^{\prime}|\mc{E}_e (0) \rangle = e^{-\frac{i}{\hbar} f_{g}(t)} |g (t) \rangle\otimes|\mc{E}_g (0) \rangle+  e^{-\frac{i}{\hbar} f_{e}(t)} |e (t) \rangle\otimes|\mc{E}_e (0) \rangle.
\end{equation}
Projecting the environment $\mc{E}$ into the state ``$|\mc{E}_g (0) \rangle$'', we have
\begin{equation}
U |g (0) \rangle \langle \mc{E}_g (0)|U^{\prime}|\mc{E}_g (0) \rangle + U |e (0) \rangle \langle \mc{E}_g (0)|U^{\prime}|\mc{E}_e (0) \rangle = e^{-\frac{i}{\hbar} f_{g}(t)} |g (t) \rangle,
\end{equation}
equivalently written as
\begin{equation}
(U^{\prime}_{g,g}) U |g (0) \rangle + (U^{\prime}_{g,e}) U |e (0) \rangle = e^{-\frac{i}{\hbar} f_{g}(t)} |g (t) \rangle,
\end{equation}
which implies
\begin{equation}
(U^{\prime}_{g,g})  |g (0) \rangle + (U^{\prime}_{g,e})  |e (0) \rangle = e^{-\frac{i}{\hbar} f_{g}(t)} U^{\dagger}|g (t) \rangle.
\end{equation}
Therefore,
\begin{equation}
U^{\prime}_{g,g}= e^{-\frac{i}{\hbar} f_{g}(t)} \langle g (0)|U^{\dagger}|g (t) \rangle, \  \text{and} \ \ U^{\prime}_{g,e}= e^{-\frac{i}{\hbar} f_{g}(t)} \langle e (0)|U^{\dagger}|g (t) \rangle.
\end{equation}
Additionally, by projecting $\mc{E}$ into the state ``$|\mc{E}_e (0) \rangle$'' we get
\begin{equation}
	U^{\prime}_{e,g}= e^{-\frac{i}{\hbar} f_{e}(t)} \langle g (0)|U^{\dagger}|e (t) \rangle, \  \text{and} \ \ U^{\prime}_{e,e}= e^{-\frac{i}{\hbar} f_{e}(t)} \langle e (0)|U^{\dagger}|e (t) \rangle.
\end{equation}
It is straightforward to check that the operator $U^{\prime}$, which reads
\begin{equation}
	\label{Uprime}
	U^{\prime}=
	\begin{pmatrix}
	U^{\prime}_{g,g} & U^{\prime}_{g,e}  \\
		U^{\prime}_{e,g}  & U^{\prime}_{e,e}
	\end{pmatrix},
\end{equation} 
is indeed a unitary on $\mc{E}$.

In conclusion, we have constructed a unique unitary map that acts only on $\mc{E}$, but counteracts transitions in $\mc{S}$. Note that coupling the system and environment implies that the state of the system is no longer described by a wave function. Hence the usual counterdiabatic scheme evolves the density matrix $\rho_{\mc{S}}(0)$ to another density $\rho_{\mc{S}}(t)$, such that both matrices have the same populations and coherence in the instantaneous eigenbasis of $H_{0}(t)$ (which is what EASTA accomplishes, as well).

\subsection{N-level environment $\mc{E}$}

We can easily generalize the two-level analysis to an $N$-level environment. Similar to the above description, coupling the system to the environment leads to a branching state of the form
\begin{equation}
	| \psi_{\mc{SE}}(0) \rangle= \frac{1}{\sqrt{N}} \sum^{N-1}_{n=0} |n (0) \rangle\otimes|\mc{E}_n (0) \rangle,
\end{equation}
where the states $\{|\mc{E}_n (0) \rangle\}_n$ form a basis on the environment $\mc{E}$. We can then construct a unique unitary $U^{\prime}$ such that the system evolves through the adiabatic manifold in each branch of the wave function,
\begin{equation}
	(\exists! \ U^{\prime}); \ \ (U \otimes U^{\prime})| \psi_{\mc{SE}}(0) \rangle= (U_{\text{\tiny CD}} \otimes \mathds{I}_{\mc{E}})| \psi_{\mc{SE}}(0) \rangle.
\end{equation}
The proof follows the exact same strategy as the two-level case, and we find
\begin{equation}
(\forall \ (m,n) \in \llbracket 0, N-1 \rrbracket^{2} ); \ \ U^{\prime}_{m,n}= e^{-\frac{i}{\hbar} f_{m}(t)} \langle n(0)|U^{\dagger}|m(t) \rangle.
\label{mainr1}
\end{equation}
The above expression of the elements of the unitary $U^{\prime}$ is our main result, which holds for any driving $H_{0}(t)$ and any $N$-dimensional system. 

\subsection{Process cost}

Having established the general analytic form of the unitary applied on the environment, the next logical step is to compute and compare the cost of both schemes: (a) the usual counterdiabatic scheme and (b) the environment-assisted shortcut scheme presented above (cf. Figure~\ref{sta}). More specifically, we now compare the time integral of the instantaneous cost~\cite{cost5} for both driving schemes~\cite{cost1,cost2,cost3,cost4,cost5}, (a) $C_{\text{\tiny CD}}(t)= (1/\tau)\int_{0}^{t} \| H_{\text{\tiny CD}}(s) \| ds$ and (b) $C_{\text{env}}(t)= (1/\tau)\int_{0}^{t} \| H_{\text{env}}(s) \| ds$ ($\|.\|$ is the operator norm), where the driving Hamiltonian on the environment can be determined from the expression of $U^{\prime}(t)$, $H_{\text{env}}(t)= i\hbar \frac{dU^{\prime}(t)}{dt} U^{\prime \dagger}(t)$.
 
In fact, from~\myeqref{mainr1} it is not too hard to see that the field applied on the environment $H_{\text{env}}(t)$ has the same eigenvalues as the counterdiabatic field $H_{\text{\tiny CD}}(t)$, since there exists a similarity transformation between $H_{\text{env}}(t)$ and $H^{*}_{\text{\tiny CD}}(t)$. Therefore, the cost of both processes is exactly the same, $C_{\text{\tiny CD}}=C_{\text{env}}$, for any arbitrary driving $H_{0}(t)$. Details of the derivation can be found in Appendix~\ref{aa}. Note that for $t=\tau$, the above definition of the cost becomes the total cost for the duration ``$\tau$'' of the process. 

\begin{figure}[h!]
	\centering
	\subfigure[]{
		\includegraphics[width=.5\textwidth]{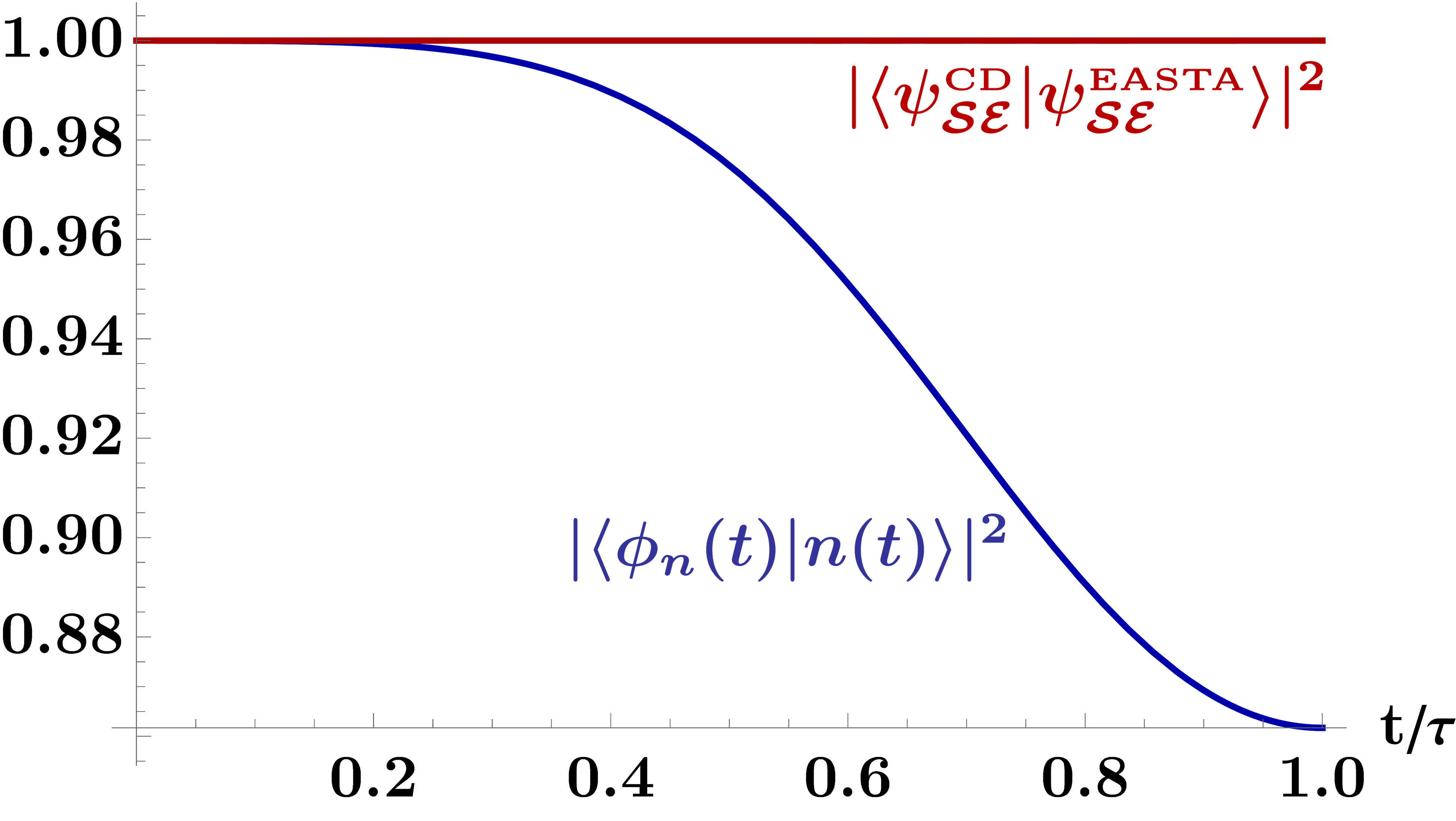}
	}
	\subfigure[]{
		\includegraphics[width=.485\textwidth]{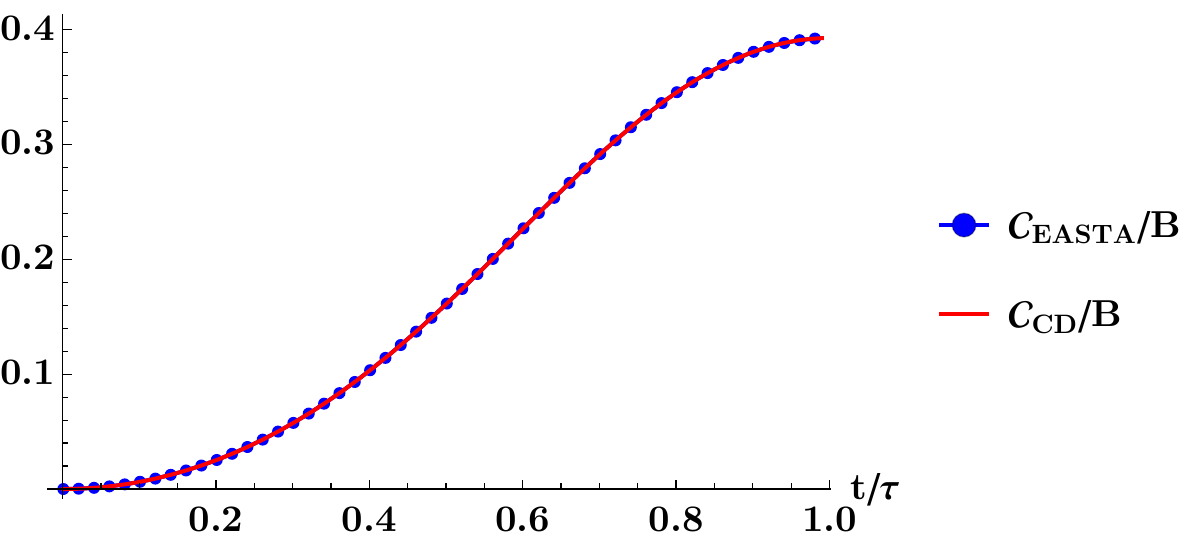}
	}
	\subfigure[]{
	\includegraphics[width=.485\textwidth]{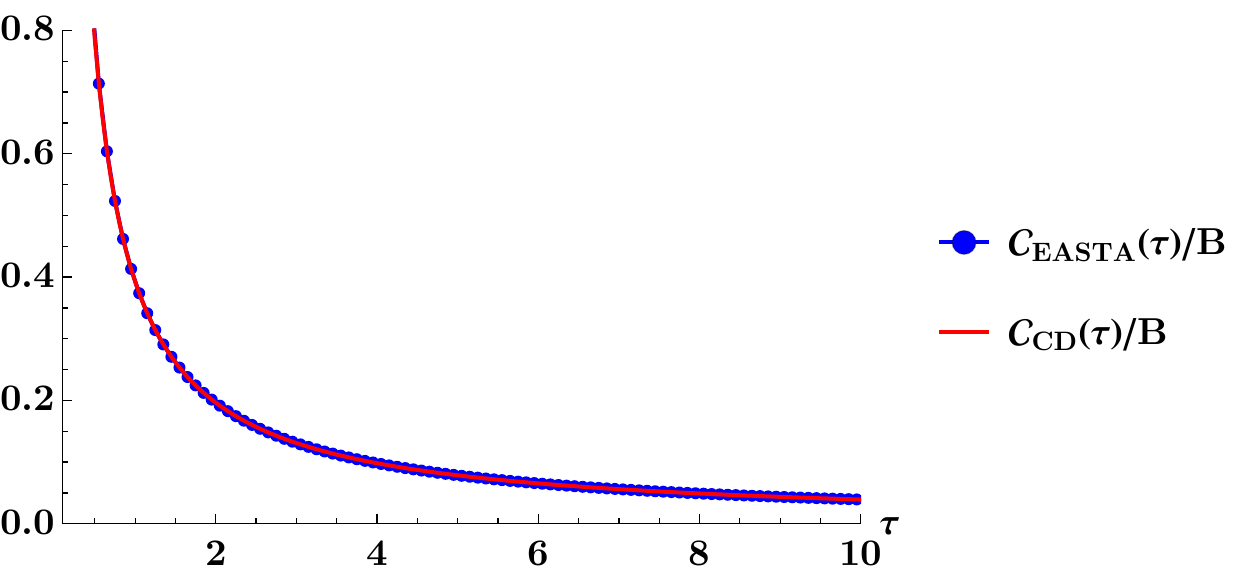}
}
	\caption[]{\label{illust1} In panel (a), the blue curve illustrates the overlap between the $n$th evolved state $| \phi_{n}(t) \rangle$ and the $n$th  instantaneous eigenstate $|n(t) \rangle$ of the Hamiltonian $H_{0}(t)$. This curve shows that the driving $H_{0}(t)$ evolves the system beyond the adiabatic manifold. The red curve illustrates that EASTA guarantees an exact evolution through the adiabatic manifold. In panel (b), we illustrate the cost of both the EASTA and the CD schemes, and numerically verify that they are equal $\mc{C}=C_{\text{\tiny CD}}=C_{\text{env}}$ for all $t/\tau \in [0,1]$, and $\tau=1$. Note that in the illustrations we pick the driving field $J(t) = B \cos^{2}\left(\frac{\pi t}{2\tau}\right)$ and $B=1$. In panel (c), we illustrate the costs for different values of $\tau$. For infinitely fast processes ($\tau \rightarrow 0$) the cost diverges and it tends to zero for infinitely slow processes ($\tau \rightarrow \infty$).}
\end{figure}

\subsection{Illustration}

We illustrate our results in a simple two-qubit model, where the system and and the environment are each described by a single qubit. Note that the environment can live in a larger Hilbert space while still characterized as a virtual qubit \cite{QD20}. The aforementioned virtual qubit notion simply means that the state of the environment is of rank equal to two.

We choose a driving Hamiltonian $H_{0}(t)$, such that
\begin{equation}
H_{0}(t)= \frac{B}{2} \sigma_x + \frac{J(t)}{2} \sigma_z,
\end{equation}
where $J(t)$ is the driving/control field, $B$ is a constant, $\sigma_z$ and $\sigma_x$ are Pauli matrices. Depending on the physical context, $B$ and $J(t)$ can be interpreted in various ways. In particular, as noted in Ref.~\cite{analyticqubit}, in some contexts the constant $B$ can be regarded as the energy splitting between the two levels~\cite{h1,h2,h3}, and in others, the driving $J(t)$ can be interpreted as a time-varying energy splitting between the states~\cite{J1,J2,J3,J4}. To illustrate our results we choose 
\begin{equation}
(\forall \ t \in [0,\ \tau]); \ J(t) = B \cos^{2}\left(\frac{\pi t}{2\tau}\right).
\end{equation}
The above driving evolves the system beyond the adiabatic manifold, and we quantify this by plotting, in Figure~\ref{illust1}, the overlap between the evolved state $| \phi_{n}(t) \rangle \equiv U(t) |n(0) \rangle$ and the instantaneous eigenstate $|n(t) \rangle$ of the Hamiltonian $H_{0}(t)$, for $n \in \{g,e\}$. To illustrate our main result, we also plot the overlap between the states resulting from the two shortcut schemes (illustrated in Figure~\ref{sta}): the first scheme is the usual counterdiabatic (CD) driving where we add a counterdiabatic field $H_{\text{\tiny CD}}$ to the system of interest, and we note the resulting composite state as ``$|\psi^{\text{\tiny CD}}_{\mc{SE}} \rangle$''. The second scheme is the environment-assisted shortcut to adiabaticity (EASTA), and we note the resulting composite state as ``$|\psi^{\text{\tiny EASTA}}_{\mc{SE}} \rangle$''. Confirming our analytic results, the local driving on the environment ensures that the system evolves through the adiabatic manifold at all times, since the state overlap is equal to one for all $t \in [0,\tau]$.

Finally, we compute and plot the cost of both shortcut schemes and verify that they are both equal to each other for all times ``$t$'' (cf. Figure~\ref{illust1}, panel (b)), and for all ``$\tau$'' (cf. Figure~\ref{illust1}, panel (c)).

\section{Concluding remarks}

\subsection{Summary}

In the present manuscript, we considered branching states $\{|\psi_{\mc{SE}}\rangle\}$, on a Hilbert space $\mc{H}_{\mc{S}}\otimes \mc{H}_{\mc{E}}$ of arbitrary dimension, and we derived the general analytic form of the time-dependent driving on $\mc{H}_{\mc{E}}$ which guarantees that the system $\mc{S}$ evolves through the adiabatic manifold at all times. Through this \textit{Environment-Assisted Shortcuts To Adiabaticity} scheme, we explicitly showed that the environment can act as a proxy to control the dynamics of the system of interest. Moreover, for branching states $|\psi_{\mc{SE}}\rangle$ with equal branch probabilities, we further proved that the cost associated with the EASTA technique is exactly equal to that of counterdiabatic driving. We illustrated our results in a simple two-qubit model, where the system and the environment are each described by a single qubit.

It is interesting to note, that while we focused in the present manuscript on counterdiabatic driving, the technique can readily be generalized to any type of control unitary map ``$U_{\text{\tiny control}}$'', resulting in a desired evolved state $|\kappa_n(t) \rangle \equiv U_{\text{\tiny control}} |n(0) \rangle$. The corresponding unitary $U^{\prime}$ on $\mc{H}_{\mc{E}}$ has then the form
\begin{equation}
(\forall \ (m,n) \in \llbracket 0, N-1 \rrbracket^{2} ); \ \ U^{\prime}_{m,n}= \langle n(0)|U^{\dagger}|\kappa_m(t) \rangle.
\label{mainr111}
\end{equation}
In the special case, for which the evolved state is equal to the $n$th instantaneous eigenstate of $H_{0}(t)$ (with a phase factor),
\begin{equation}
|\kappa_n(t) \rangle=e^{-\frac{i}{\hbar} f_{n}(t)} |n(t) \rangle,
\end{equation}
we recover the main result of the manuscript. The above generalization illustrates the broad scope of our results. Any control unitary on the system $\mc{S}$ can be realized solely by acting on the environment $\mc{E}$, without altering the dynamics of the system of interest $\mc{S}$ (i.e., for any arbitrary driving $H_{0}(t)$, hence any driving rate).

\subsection{Envariance and pointer states}

In the present work, we leveraged the presence of an environment to induce desired dynamics in a quantum system.  Interestingly, our novel method for shortcuts to adiabaticity relies on branching states, which play an essential role in decoherence theory and in the framework of Quantum Darwinism.

In open system dynamics~\cite{decoh1,decoh2,decoh3}, the interaction between system and environment superselects states that survive the decoherence process, aka the pointer states~\cite{pointer1,pointer2}.  It is exactly these pointer states that are the starting point of our analysis, and for which EASTA is designed.  While previous studies \cite{psta6,open1,open2} have explored STA methods for open quantum systems, to the best of our understanding, the environment was only considered as a passive source of additional noise described by quantum master equations. In our paradigm, we recognize the active role that an environment plays in quantum dynamics, which is inspired by envariance and reminiscent of the mind-set of Quantum Darwinism.  In this framework~\cite{QD1,QD2,QD3,QD4,QD5,QD6,QD7,QD8,QD9,QD10,QD11,QD12,QD13,QD14,QD15,QD16,QD17,QD19,QD20},  the environment is understood as a communication channel through which we learn about the world around us, i.e., we learn about the state of systems of interest by eavesdropping on environmental degrees of freedom~\cite{QD20}. 

Thus, in true spirit of the teachings by Wojciech H. Zurek we have understood the agency of quantum environments and the useful role they can assume. To this end,  we have applied a small part of the many lessons we learned from working with Wojciech, to connect and merge tools from seemingly different areas of physics to gain a deeper and more fundamental understanding of nature.

\appendix
\numberwithin{equation}{section}
\renewcommand{\theequation}{\thesection\arabic{equation}}

\section{Cost of environment-assisted shortcuts to adiabaticity}
\label{aa}
\setcounter{equation}{0}

In this appendix, we show that CD and EASTA have the same cost.  Generally, we have
\begin{equation}
\begin{split}
H_{\text{env}}(t)&= i\hbar \frac{dU^{\prime}(t)}{dt} U^{\prime \dagger}(t),\\
&= i\hbar \sum_{i,j} \sum_{k} \frac{dU^{\prime}_{i,k}}{dt} (U^{\prime}_{j,k})^{*} |\mc{E}_{i}(0)\rangle 
\langle \mc{E}_{j}(0)|.\\
\end{split}
\end{equation}
From the main result in~\myeqref{mainr1}, we obtain
\begin{equation}
H_{\text{env}}(t)= \sum_{i,j} \left(\sum_{k} \left(\langle k(0)|i\hbar\partial_{t}U^{\dagger}|\psi_{i}(t) \rangle (U^{\prime}_{j,k})^{*} + i\hbar\langle k(0)|U^{\dagger}|\partial_{t}\psi_{i}(t) \rangle(U^{\prime}_{j,k})^{*}\right)\right)|\mc{E}_{i}(0)\rangle 
\langle \mc{E}_{j}(0)|.
\end{equation}
Given that $H_{0}(t)= i\hbar \frac{dU(t)}{dt} U^{\dagger}(t)$, we also have
\begin{equation}
H_{\text{env}}(t)= \sum_{i,j} \left(\sum_{k} \left(\langle k(0)|(-U^{\dagger}H_{0})|\psi_{i}(t) \rangle (U^{\prime}_{j,k})^{*} + i\hbar\langle k(0)|U^{\dagger}|\partial_{t}\psi_{i}(t) \rangle(U^{\prime}_{j,k})^{*}\right)\right)|\mc{E}_{i}(0)\rangle 
\langle \mc{E}_{j}(0)|,
\end{equation}
which implies
\begin{equation}
H_{\text{env}}(t)= \sum_{i,j} \left(\sum_{k} \left(\langle k(0)|(-U^{\dagger}H_{0})|\psi_{i}(t) \rangle (U^{\prime}_{j,k})^{*} + \langle k(0)|U^{\dagger}H|\psi_{i}(t) \rangle(U^{\prime}_{j,k})^{*}\right)\right)|\mc{E}_{i}(0)\rangle 
\langle \mc{E}_{j}(0)|,
\end{equation}
and hence,
\begin{equation}
H_{\text{env}}(t)= \sum_{i,j} \left(\sum_{k} \langle k(0)|U^{\dagger}H_{\text{\tiny CD}}|\psi_{i}(t) \rangle(U^{\prime}_{j,k})^{*}\right)|\mc{E}_{i}(0)\rangle 
\langle \mc{E}_{j}(0)|.
\end{equation}
Using $| \phi_{k}(t) \rangle \equiv U(t) |k(0) \rangle$ we can write
\begin{equation}
H_{\text{env}}(t)= \sum_{i,j} \left(\sum_{k}  \langle \psi_{j}(t)|\phi_k(t) \rangle \langle \phi_k(t)|H_{\text{\tiny CD}}|\psi_{i}(t) \rangle\right)|\mc{E}_{i}(0)\rangle 
\langle \mc{E}_{j}(0)|.
\end{equation}
Therefore,
\begin{equation}
H_{\text{env}}(t)= \sum_{i,j} \left(  \langle \psi_{j}(t)|H_{\text{\tiny CD}}|\psi_{i}(t) \rangle\right)|\mc{E}_{i}(0)\rangle 
\langle \mc{E}_{j}(0)|.
\end{equation}
By definition, we also have
\begin{equation}
H_{\text{\tiny CD}}= \sum_{i,j} \left(  \langle \psi_{i}(t)|H_{\text{\tiny CD}}|\psi_{j}(t) \rangle\right)|\psi_{i}(t)\rangle 
\langle \psi_{j}(t)|,
\end{equation}
hence,
\begin{equation}
H^{T}_{\text{\tiny CD}}=H^{*}_{\text{\tiny CD}}= \sum_{i,j} \left(  \langle \psi_{j}(t)|H_{\text{\tiny CD}}|\psi_{i}(t) \rangle\right)|\psi_{i}(t)\rangle 
\langle \psi_{j}(t)|.
\end{equation}
Thus, there exists a similarity transformation between $H^{*}_{\text{\tiny CD}}$ and $H_{\text{env}}$,  and $C_{\text{\tiny CD}}=C_{\text{env}}$ for any arbitrary driving $H_{0}(t)$. The similarity transformation is given by the matrix $S=\sum_{j} |\mc{E}_j(0)\rangle \langle \psi_j(t)|$, such that $SH^{*}_{\text{\tiny CD}}S^{-1}=H_{\text{env}}$. Since we proved that the Hamiltonians $H_{\text{\tiny CD}}$ and $H_{\text{env}}$ have the same eigenvalues, our result can be valid for other definitions of the cost function $\mc{C}$ which might involve other norms (e.g., the Frobenius norm).

\section{Generalization to arbitrary branching probabilities}
\label{a}

Finally we briefly inspect the case of non-even branching states.  We begin by noting the consequences of our assumptions. In particular, we have assumed that the state of system+environment evolves unitarly. Thus,  consider a joint map of the form $U \otimes M$, where $U$ is a unitary on $\mc{S}$. Then, it is a simple exercise to show that the map $M$, on $\mc{E}$, is also unitary, $MM^{\dagger}=M^{\dagger}M=\mathbb{I}$. In what follows, we prove by contradiction that there exists no unitary map $M$ that suppresses transitions in $\mc{S}$, for branching states with arbitrary probabilities.

Consider
\begin{equation}
| \psi_{\mc{SE}}(0) \rangle=  \sum^{N-1}_{n=0} \sqrt{p_n} |n (0) \rangle \bigotimes^{N_{\mc{E}}}_{l=1}|\mc{E}^{l}_n (0) \rangle,
\label{bran1}
\end{equation}
and assume that there exists a unitary map $M$ on $\mc{E}$ that suppresses transitions in $\mc{S}$, i.e.,
\begin{equation}
\sum^{N-1}_{n=0} \sqrt{p_n} U|n (0) \rangle \otimes \left(M \bigotimes^{N_{\mc{E}}}_{l=1} |\mc{E}^{l}_n (0) \rangle \right)=\sum^{N-1}_{n=0} \sqrt{p_n} e^{-\frac{i}{\hbar} f_{n}(t)}  |n(t) \rangle \otimes \left(\bigotimes^{N_{\mc{E}}}_{l=1} |\mc{E}^{l}_n (0) \rangle \right).
\end{equation}
Following the same steps of Section~\ref{sec3} we obtain
\begin{equation}
(\forall \ (m,n) \in \llbracket 0, N-1 \rrbracket^{2} ); \ \ M_{m,n}= \sqrt{\frac{p_m}{p_n}} e^{-\frac{i}{\hbar} f_{m}(t)} \langle n(0)|U^{\dagger}|m(t) \rangle.
\label{map1}
\end{equation}
Comparing the above map with our main result in~\myeqref{mainr1}, we conclude that the additional factor $\sqrt{\frac{p_m}{p_n}}$ violates unitarity,  and hence we conclude that EASTA cannot work for non-even branching states \eqref{bran1}.

This can be seen more explicitly from the form of the matrices $MM^{\dagger}$ and $M^{\dagger}M$. Generally, and by dropping the superscript in environmental states $\bigotimes^{N_{\mc{E}}}_{l=1}|\mc{E}^{l}_n (0) \rangle \equiv |\mc{E}_n (0) \rangle$, we have
\begin{equation}
MM^{\dagger}= \sum_{i,j,k} M_{i,k} M^{*}_{j,k} |\mc{E}_i(0)\rangle \langle \mc{E}_j(0)|,
\end{equation}
from the expression of the elements of $M$ (cf.~\myeqref{map1}), and by adopting the notation $| \phi_{n} \rangle \equiv U(t) |n(0) \rangle$, we get
\begin{equation}
MM^{\dagger}= \sum_{i,j,k} \frac{\sqrt{p_ip_j}}{p_k} \langle \phi_k | \psi_i \rangle \langle \psi_j | \phi_k \rangle|\mc{E}_i(0)\rangle \langle \mc{E}_j(0)|,
\end{equation}
which implies
\begin{equation}
MM^{\dagger}= \mathbb{I}+\sum_{i,j} \sqrt{\frac{p_j}{p_i}} \langle \psi_j | D_{(i)}|\psi_i  \rangle|\mc{E}_i(0)\rangle \langle \mc{E}_j(0)|,
\label{counter1}
\end{equation}
such that the matrix
\begin{equation}
D_{(i)} = \sum_k \frac{p_i}{p_k} |\phi_k\rangle \langle \phi_k |-\mathbb{I}
\end{equation}
is diagonal in the basis spanned by the orthonormal vectors $\{|\phi_k\rangle\}_k$. This matrix is generally (for any choice of $H_{0}(t)$ and initial state of $\mc{S}$) different from the null matrix for non-equal branch probabilities. A similar decomposition can be made for the matrix $M^{\dagger}M$, such that
\begin{equation}
M^{\dagger}M= \mathbb{I}+\sum_{i,j} \sqrt{\frac{p_i}{p_j}} \langle \phi_j | \mc{D}_{(i)}|\phi_i  \rangle|\mc{E}_i(0)\rangle \langle \mc{E}_j(0)|,
\label{counter2}
\end{equation}
where
\begin{equation}
\mc{D}_{(i)} = \sum_k \frac{p_k}{p_i} |\psi_k\rangle \langle \psi_k |-\mathbb{I}.
\end{equation}

In conclusion, for branching states with non-equal probabilities there is no unitary map that guarantees that the system evolves through the adiabatic manifold at all times and for any arbitrary driving $H_{0}(t)$. Hence we can realize the EASTA technique only for a system maximally entangled with its environment (cf.~\myeqref{bran1} with $\sqrt{p_n}=1/\sqrt{N}$ for all $n \in \llbracket 0, \ N-1 \rrbracket$), or in the general case (non-equal branch probabilities) when we can access an extended Hilbert space.

\vspace{6pt} 

\funding{S.D. acknowledges support from the U.S. National Science Foundation under Grant No. DMR-2010127. This research was supported by grant number FQXi-RFP-1808 from the Foundational Questions Institute and Fetzer Franklin Fund, a donor advised fund of Silicon Valley Community Foundation (SD).}

\acknowledgments{We would like to thank Wojciech H. Zurek for many years of mentorship and  his unwavering patience and willingness to teach us how to think about the mysteries of the quantum universe. Enlightening discussions with Agniva Roychowdhury are gratefully acknowledged.}

\conflictsofinterest{The authors declare no conflict of interest.} 

\bibliography{opm}

\end{document}